\documentclass[showpacs]{revtex4} 
\usepackage{epsfig,amsmath,amssymb,graphicx}
\usepackage{natbib}
\usepackage[english]{babel} 

\begin{document}
\vspace{0mm}
\title{Accounting for the long-range forces in the model of hard spheres} %
\author{Yu.M. Poluektov}
\email{yuripoluektov@kipt.kharkov.ua} %
\affiliation{National Science Center ``Kharkov Institute of Physics
and Technology'', 61108 Kharkov, Ukraine}

\begin{abstract}
A method for taking into account the long-range potential of atoms
in the framework of the hard-sphere model is proposed. It is shown
that thermodynamic quantities can be represented as a sum of three
contributions -- that of an ideal gas, the interaction of hard
spheres, and the long-range potential. In the leading approximation
on density, the corrections to the virial coefficient and heat
capacity due to the smooth component of the potential are
calculated. Attention is drawn to the fact that the effects
determined by the long-range part of the potential can be described
in the framework of the self-consistent field model.
\newline%
\end{abstract}
\maketitle

\section{Introduction}\vspace{-0mm} 

In the theory of classical many-particle systems, an important role
is played by the model of hard spheres
\cite{GKB,Temperley,Croxton,Santos,Fisher}. Although it is not
possible to calculate the configuration integral even in the case of
such a simple pair interaction potential,
 a sufficiently large number of virial coefficients are known here for
low-density systems. The first four coefficients are known exactly,
and higher order coefficients are calculated by numerical methods
and the Monte Carlo method
\cite{GKB,Temperley,Croxton,Santos,Fisher}. Within the framework of
the hard-sphere model, there is also an exact solution of the
Percus-Yevick integral equation for the pair correlation function
\cite{Wertheim1,Wertheim2}. A significant drawback of the
hard-sphere model is that the configuration integral and virial
coefficients for it do not depend on temperature.

In more realistic pair potentials, which depend only on the distance
between particles, as a rule, it is possible to distinguish a region
of strong repulsion at small distances and a rather smoothly varying
part of the potential at large distances. The repulsive part usually
differs little from the potential of hard spheres, so it is natural
to model it by the potential of a hard sphere and, along with this,
take into account the contribution to thermodynamic quantities of
the long-range part of the potential.

In this work, based on such a decomposition of the pair potential, a
method is proposed for calculating the thermodynamic characteristics
of a gas and a liquid. It is shown that the free energy and
thermodynamic quantities such as the pressure, entropy, heat
capacity, chemical potential can be represented as a sum of three
contributions -- that of an ideal gas, the interaction of hard
spheres, and the long-range potential. Corrections for the
long-range part of the potential to thermodynamic quantities, in
particular to the virial coefficient and heat capacity, are
calculated in the leading approximation on density. It is noted that
the self-consistent field model is applicable to describe the
long-range interaction.

\section{Thermodynamic relations in the model of hard spheres with account of
the long-range part of the potential }\vspace{-0mm} The free energy
of a system of $N$ particles $F=-T\ln Z_N$ is
calculated through the partition function, which can be represented as %
\begin{equation} \label{01}
\begin{array}{l}
\displaystyle{%
  Z_N=\left(\frac{r_0}{\Lambda}\right)^{\!3N}\frac{Q_N}{N!},
}
\end{array}
\end{equation}
where
\begin{equation} \label{02}
\begin{array}{l}
\displaystyle{%
  \Lambda\equiv\left(\frac{h^2}{2\pi mT}\right)^{\!1\!/2}
}
\end{array}
\end{equation}
is the thermal de Broglie wavelength, $h$ is the Planck's constant,
$T$ is the temperature, $m$ is the mass of an atom. The
configuration integral in (\ref{01}) is defined by the formula
\begin{equation} \label{03}
\begin{array}{l}
\displaystyle{%
  Q_N=\frac{1}{r_0^{3N}}\int e^{-\beta U(q)}dq,
}
\end{array}
\end{equation}
where the designations are used: $q\equiv\{{\bf r}_1,{\bf
r}_2,\ldots,{\bf r}_N\}, dq\equiv d{\bf r}_1d{\bf r}_2\ldots d{\bf
r}_N$ and $\beta=1/T$. In formulas (\ref{01}) and (\ref{03}), a
certain characteristic distance $r_0$ is introduced, which
determines the conditional size of an atom. In what follows, this
parameter will signify the radius of a hard sphere. The interaction
between particles is realized through the pair potential, so that in (\ref{03}) %
\begin{equation} \label{04}
\begin{array}{l}
\displaystyle{%
  U(q)\equiv\sum_{N\geq i>j\geq 1}U(r_{ij}),
}
\end{array}
\end{equation}
where for brevity $r_{ij}\equiv|{\bf r}_i-{\bf r}_j|$. Then the
exponent in (\ref{03}) can be written in the form
$\displaystyle{ e^{-\beta U(q)}=\prod_{N\geq i>j\geq 1}e^{-\beta U(r_{ij})}}$. %
We choose the potential of the pairwise interaction of atoms as a
sum of the potential of hard spheres $U_H(r_{ij})$ and the
long-range part $U_L(r_{ij})$: %
\begin{equation} \label{05}
\begin{array}{l}
\displaystyle{%
  U(r_{ij})\equiv U_H(r_{ij})+U_L(r_{ij}),
}
\end{array}
\end{equation}
where
\begin{equation} \label{06}
\begin{array}{l}
\displaystyle{%
  U_H(r_{ij}) = \left\{
               \begin{array}{l}
                 \infty,\quad r<r_0, \vspace{2mm} \\ %
                 0, \hspace{5.3mm} r>r_0.              %
               \end{array} \right. %
}
\end{array}
\end{equation}
The structure of the form (\ref{05}) is inherent, for example, for
the Sutherland potential
\begin{equation} \label{07}
\begin{array}{l}
\displaystyle{%
  U(r_{ij}) = \left\{
               \begin{array}{l}
                 \hspace{2mm}\infty, \hspace{16mm} r_{ij}<r_0,                                         \vspace{2mm} \\ %
                 \displaystyle{
                 -\varepsilon\left(\frac{r_0}{r_{ij}}\right)^{\!6}, \hspace{5mm} r_{ij}>r_0. }             %
               \end{array} \right. %
}
\end{array}
\end{equation}
A similar form is characteristic for other model potentials. So for
the modified Lennard-Jones potential we have
\begin{equation} \label{08}
\begin{array}{l}
\displaystyle{%
  U(r_{ij}) = \left\{
               \begin{array}{l}
                 \hspace{8mm}\infty, \hspace{30.3mm} r_{ij}<r_0,                                         \vspace{2mm} \\ %
                 \displaystyle{
                 4\varepsilon\!\left[\left(\frac{r_0}{r_{ij}}\right)^{\!12}-\,\left(\frac{r_0}{r_{ij}}\right)^{\!6}\right], \hspace{5mm} r_{ij}>r_0. }             %
               \end{array} \right. %
}
\end{array}
\end{equation}

The configuration integral of the model of hard spheres has the form %
\begin{equation} \label{09}
\begin{array}{l}
\displaystyle{%
  Q_N^{(H)}= \frac{1}{r_0^{3N}}\int\!\prod_{N\geq i>j\geq 1}\theta\left(\frac{r_{ij}}{r_0}-1\right)dq, %
}
\end{array}
\end{equation}
where
\begin{equation} \label{10}
\begin{array}{l}
\displaystyle{%
  \theta(x) = \left\{
               \begin{array}{l}
                 1,\quad x>0, \vspace{2mm} \\ %
                 0, \quad x<0,               %
               \end{array} \right. %
}
\end{array}
\end{equation}
is the stepwise function. With such a decomposition, the full
configuration integral can be represented as a sum
\begin{equation} \label{11}
\begin{array}{l}
\displaystyle{%
  Q_N=Q_N^{(H)}+Q_N^{(L)}, %
}
\end{array}
\end{equation}
where
\begin{equation} \label{12}
\begin{array}{l}
\displaystyle{%
  Q_N^{(L)}= \frac{1}{r_0^{3N}}\int\!\prod_{N\geq i>j\geq 1}\theta\left(\frac{r_{ij}}{r_0}-1\right)\Bigg[\prod_{N\geq k>r\geq 1}e^{-\beta U_L(r_{kr})}-1\Bigg]dq. %
}
\end{array}
\end{equation}
Thus, for potentials that have the form of a sum of the potential of
hard spheres and the smooth long-range part, the configuration
integral also has the form of a sum of contributions from the
potential of hard spheres and the long-range part (\ref{11}). It is
convenient to introduce the ``reduced'' configuration integrals
\begin{equation} \label{13}
\begin{array}{l}
\displaystyle{%
  \tilde{Q}_N^{(H)}\equiv\frac{Q_N^{(H)}}{Q_N^{(0)}}, \qquad \tilde{Q}_N^{(L)}\equiv\frac{Q_N^{(L)}}{Q_N^{(0)}}, %
}
\end{array}
\end{equation}
where
$\displaystyle{Q_N^{(0)}\equiv\left(\frac{V}{r_0^3}\right)^{\!N}}$
is the configuration integral of an ideal gas. Then the free energy
can be written as a sum of three contributions $F=F_0+F_H+F_L$, where %
\begin{equation} \label{14}
\begin{array}{l}
\displaystyle{%
  F_0=-NT\ln \left(\frac{e\upsilon}{\Lambda^3}\right) %
}
\end{array}
\end{equation}
is the free energy of an ideal gas,
$\displaystyle{\upsilon=\frac{V}{N}}$ is the volume per one particle, %
\begin{equation} \label{15}
\begin{array}{l}
\displaystyle{%
  F_H=-T\ln\tilde{Q}_N^{(H)} %
}
\end{array}
\end{equation}
is the contribution from collisions of hard spheres,
\begin{equation} \label{16}
\begin{array}{l}
\displaystyle{%
  F_L=-T\ln\left(1+\frac{\tilde{Q}_N^{(L)}}{\tilde{Q}_N^{(H)}}\right) %
}
\end{array}
\end{equation}
is the contribution from the long-range part of the interaction.
Note that not only $\tilde{Q}_N^{(L)}$, but also the configuration
integral of the model of hard spheres $\tilde{Q}_N^{(H)}$ enters
into $F_L$. Other thermodynamic quantities can also be represented
as a sum of three contributions. So far, no approximations have been
made in deriving the formulas.

\section{Calculation of corrections for the long-range part of the potential}\vspace{-0mm}
In a sufficiently dilute system, it is possible to account for the
interaction using the group expansion in powers of density
\cite{GKB,Temperley,Croxton,Santos,Fisher}. For this purpose, the
transition to Mayer functions
\begin{equation} \label{17}
\begin{array}{l}
\displaystyle{%
  f_H(r_{ij})=\theta\left(\frac{r_{ij}}{r_0}-1\right)-1, \qquad f_L(r_{ij})=e^{-\beta U(r_{ij})}-1%
}
\end{array}
\end{equation}
is employed. Taking into account the first correction for the
dimensionless density $n\upsilon_a=\upsilon_a/\upsilon$, where
$\displaystyle{\upsilon_a=\frac{4\pi}{3}r_0^3}$ is the ``volume'' of an atom, %
for the configuration integral of the model of hard spheres we have
\begin{equation} \label{18}
\begin{array}{l}
\displaystyle{%
  \tilde{Q}_N^{(H)}\approx 1-\frac{N}{2}\frac{\upsilon_a}{\upsilon}. %
}
\end{array}
\end{equation}
Note that the use of the formula $\ln(1+x)\approx x$ in calculating
the free energy (\ref{15}) is valid under the condition
$N<\upsilon/\upsilon_a$ that is satisfied for a highly dilute gas \cite{Croxton}. %

In the leading approximation on density, the configuration integral
for the long-range part of the potential is given by the formula
\begin{equation} \label{19}
\begin{array}{l}
\displaystyle{%
  \tilde{Q}_N^{(L)}=\frac{N(N-1)}{2}\frac{4\pi}{V}\int_{r_0}^\infty\! dr r^2 \Big[e^{-\beta U_L(r)}-1\Big]. %
}
\end{array}
\end{equation}
Taking into account the main correction to formulas of the theory of
an ideal gas in the ratio $\upsilon_a/\upsilon$, the full free
energy takes the form
\begin{equation} \label{20}
\begin{array}{l}
\displaystyle{%
  F=-NT\ln \left(\frac{e\upsilon}{\Lambda^3}\right)+\frac{NT}{2}\frac{\upsilon_a}{\upsilon}-\frac{3}{2}NT\frac{\upsilon_a}{\upsilon}J(T), %
}
\end{array}
\end{equation}
where
\begin{equation} \label{21}
\begin{array}{l}
\displaystyle{%
  J(T)=\int_{1}^\infty\! dx x^2 \Big[e^{-\beta U_L(r_0x)}-1\Big]. %
}
\end{array}
\end{equation}
Let us present formulas for the basic thermodynamic quantities in
this approximation. The gas equation of state has the form
\begin{equation} \label{22}
\begin{array}{l}
\displaystyle{%
  P=\frac{T}{\upsilon}\left(1+\frac{B}{\upsilon}\right). %
}
\end{array}
\end{equation}
The virial coefficient $B$ in the model of hard spheres is
independent of temperature $B_H=\upsilon_a/2$. The calculation of
the virial coefficient with account of the long-range part of the
potential based on the formulas (\ref{16}),\,(\ref{17}),\,(\ref{21}) gives %
\begin{equation} \label{23}
\begin{array}{l}
\displaystyle{%
  B(T)=B_H\big[1-3J(T)\big]. %
}
\end{array}
\end{equation}
For example, for the Sutherland potential (\ref{07}), the integral
(\ref{21}) has the form
\begin{equation} \label{24}
\begin{array}{l}
\displaystyle{%
  J(T)=\int_{1}^\infty\! dx x^2 \Big[e^{\,\varepsilon/Tx^6}-1\Big]=\frac{1}{6}\sqrt{\frac{\varepsilon}{T}}\int_{0}^{\varepsilon/T}\!\!\!\frac{dy}{y^{3/2}}\big(e^y-1\big). %
}
\end{array}
\end{equation}
Accounting for the long-range component of the potential leads to
the fact that at low temperatures the virial coefficient becomes
negative, changing sign at the Boyle temperature $T_B$, which is
determined by the formula
\begin{equation} \label{25}
\begin{array}{l}
\displaystyle{%
  J(T_B)=\frac{1}{3}. %
}
\end{array}
\end{equation}
In the case of the Sutherland potential $T_B=1.17\varepsilon$.

For the entropy $S=-\big(\partial F/\partial T\big)_{V,N}$ in this
approximation, we have
\begin{equation} \label{26}
\begin{array}{l}
\displaystyle{%
  S=N\ln \left(\frac{\upsilon e^{5/2}}{\Lambda^3}\right)-\frac{N}{2}\frac{\upsilon_a}{\upsilon}+\frac{3N}{2}\frac{\upsilon_a}{\upsilon}\left(J+T\frac{dJ}{dT}\right). %
}
\end{array}
\end{equation}
Note that the contribution to the entropy from collisions of hard
spheres does not depend on temperature, and the contribution to the
temperature dependence of entropy comes only from the long-range
interaction.

The energy $E=F+ST$ is determined by the formula
\begin{equation} \label{27}
\begin{array}{l}
\displaystyle{%
  E=\frac{3}{2}NT\left(1+\frac{\upsilon_a}{\upsilon}\,T\frac{dJ}{dT}\right). %
}
\end{array}
\end{equation}
Collisions of hard spheres also do not contribute to the total
energy.

Let us also give a formula for the chemical potential $\mu=\big(\partial F/\partial N\big)_{T,V}$: %
\begin{equation} \label{28}
\begin{array}{l}
\displaystyle{%
  \mu=-T\ln \left(\frac{\upsilon}{\Lambda^3}\right)+T\frac{\upsilon_a}{\upsilon}-\frac{3}{2}T\frac{\upsilon_a}{\upsilon}J. %
}
\end{array}
\end{equation}

Since the contribution to the entropy of collisions of  hard spheres
does not depend on temperature, such collisions do not contribute to
the heat capacity either. The contribution to the heat capacity is
determined only by the long-range part of the interaction potential
\begin{equation} \label{29}
\begin{array}{l}
\displaystyle{%
  C_V=\frac{3}{2}N\!\left[1+\frac{\upsilon_a}{\upsilon}\,T\!\left(2\frac{dJ}{dT}+T\frac{d^2J}{dT^2}\right)\right]. %
}
\end{array}
\end{equation}
At high temperatures $T\gg \varepsilon$ for the Sutherland potential
(\ref{07}) $\displaystyle{J=\frac{1}{3}\!\left(\frac{\varepsilon}{T}\right)+\frac{1}{18}\!\left(\frac{\varepsilon}{T}\right)^2 }$, %
so that the heat capacity with account of the main correction for
the long-range interaction takes the form
\begin{equation} \label{30}
\begin{array}{l}
\displaystyle{%
  C_V=\frac{3}{2}N\!\left(1+\frac{\upsilon_a}{\upsilon}\frac{\varepsilon^2}{9T^2}\right). %
}
\end{array}
\end{equation}
Then for energy we have
\begin{equation} \label{31}
\begin{array}{l}
\displaystyle{%
  E=\frac{3}{2}NT\!\left(1-\frac{\upsilon_a}{3\upsilon}\frac{\varepsilon}{T}-\frac{\upsilon_a}{9\upsilon}\frac{\varepsilon^2}{T^2}\right). %
}
\end{array}
\end{equation}
It is obvious that $C_V=\big(\partial E/\partial T\big)_{V,N}$.

\section{Conclusion}\vspace{-0mm}

Since the potential of repulsion of atoms at small distances is
usually known poorly, it is quite acceptable to use the model of
hard spheres to take into account the short-range correlations in
systems of many particles. Along with a short-range repulsion, the
realistic potential contains a smooth component describing the
interaction of atoms at long distances. The description of
contribution of the long-range interaction to the thermodynamic
quantities of dilute systems has been considered in this work.

The method of decomposition of the pair potential of interaction
between particles into the solid core and the long-range part,
proposed in this paper, is important in connection with the question
of possibility of using the self-consistent field method in the
theory of dense systems. The idea of using the self-consistent field
method, which allows to effectively describe phase transitions, in
the theory of solids was proposed many years ago by Vlasov
\cite{Vlasov}. Vlasov's approach was subjected to criticism by many
leading physicists. Nevertheless, the self-consistent field method,
although without firm justification, was later employed as well to
construct the statistical theory of the crystal state
\cite{Bazarov1,Bazarov2} and the thermodynamic perturbation theory
\cite{Poluektov}.

Now, apparently, we can conclude that both Vlasov and his critics
were right. Obviously, the short-range correlations of particles
cannot be described in the framework of the self-consistent field
model, but it is natural to take them into account in the framework
of the model of hard spheres. The interaction of atoms through a
smooth long-range part of the potential can now be taken into
account by the self-consistent field method. Therefore, the
decomposition of the potential into the short-range and long-range
parts and the using of the mean-field theory to take into account
the smooth part of the interaction may be important for the further
development of the theory of dense gases, liquids, and solids.

\end{document}